\documentclass[apl, letterpaper, fleqn, floatfix, showpacs, showkeys ]{revtex4}

\newcommand{\ve}{\epsilon}

\newcommand{\br}{{\bf r}}
\newcommand{\bq}{{\bf q}}

\newcommand{\FG}{\widetilde{G}}

\newcommand{\nimp}{n_i}

\newcommand{\barn}{\bar{n}}

\newcommand{\Drho}{\delta\rho}
\newcommand{\DFrho}{\delta\widetilde{\rho}}

\newcommand{\DFphi}{\delta \widetilde{V}}

\newcommand{\Dphi}{\delta V}

\usepackage{euscript, units, amsfonts, graphicx, dcolumn, fancyhdr}
\usepackage{times}
\usepackage{times, graphicx, units,mathrsfs,euscript,upgreek,amssymb,pifont}

\begin{document}

\title{Potential fluctuations in graphene due to correlated charged impurities in substrate}

\author{R. Ani\v{c}i\'{c}}
\affiliation{Department of Applied Mathematics, University of Waterloo, Waterloo, Ontario, Canada N2L 3G1}
\author{Z. L. Mi\v{s}kovi\'{c}}
\email{zmiskovi@uwaterloo.ca}
 \affiliation{Department of Applied Mathematics, and Waterloo Institute for Nanotechnology, University of Waterloo, Waterloo, Ontario,
Canada N2L 3G1}

\date{\today}



\begin{abstract}

We evaluate the autocorrelation function of the electrostatic potential in doped graphene due to nearby charged impurities. The screening of those impurities is described by a combination of the polarization function for graphene in random phase approximation with the electrostatic Green's function of the surrounding dielectrics.
Using the hard-disk model for a two-dimensional distribution of impurities, we show that large correlation lengths between impurities can give rise to anti-correlation in the electrostatic potential, in agreement with recent experiments.

\end{abstract}

\maketitle \thispagestyle{plain}

Many future applications of graphene in electronics, photonics,\cite{Avouris_2012} and biochemical sensing\cite{Allen_2010}
are based on specific properties of the low-energy excitations of its $\pi$ electrons described as Dirac fermions. Being an all-surface material enables efficient tuning of the equilibrium charge carrier density in graphene by applying an electric potential to external gates or by doping due to controlled adsorption of atoms or molecules, but it also renders graphene extremely sensitive to the chemical and structural imperfections in the surrounding materials. For example, charged impurities are ubiquitous in SiO$_2$ that is commonly used for supporting exfoliated graphene, \cite{Ishigami_2007,Romero_2008} and are found to give rise to spatial variation of the Dirac point across graphene resulting in a quite inhomogeneous distribution of its charge carriers.\cite{Martin_2008,Zhang_2009,Deshpande_2009,Deshpande_2011,Gomez_2012}
When the Fermi energy of graphene is tuned to sit at the average position of its Dirac point, such inhomogeneity in charge carriers gives rise to a system of electron-holes puddles that is responsible for the famed conductivity minimum in a nominally neutral graphene. \cite{Tan_2007,PNAS_2007,Sarma_2011}
On the other hand, even in the cases when the average equilibrium areal number density of charge carriers $\barn$ is relatively large, as is the case in graphene on SiO$_2$,\cite{Ishigami_2007,Romero_2008} the fluctuations in charge carrier density and the corresponding electrostatic potential could play important roles in, e.g., the saturation of graphene's DC conductivity \cite{Yan_2011,Li_2011} and, possibly, in the plasmon dispersion and damping of interest for applications of graphene in plasmonics.\cite{Yan_2013}

Several experimental studies were undertaken to map the charge inhomogeneity in graphene on SiO$_2$ by using local probes such as scanning single-electron transistor,
\cite{Martin_2008}
scanning tunneling microscopy (STM) and spectroscopy,
\cite{Zhang_2009, Deshpande_2009}
Coulomb blockade spectroscopy,
\cite{Deshpande_2011}
and a combination of STM with atomic force microscopy.
\cite{Gomez_2012}
The maps of sufficiently large samples may be further used to provide better understanding of the global structure of the charge inhomogeneity in graphene by analyzing the statistical properties of the associated fluctuations of the electrostatic potential. For example, the spatial dependence of the experimentally deduced autocorrelation function (ACF) of the electrostatic potential in graphene \cite{Zhang_2009,Deshpande_2011,Gomez_2012}
may provide information about the typical size of the charged patches on graphene, whereas possible changes in the sign of the ACF may point to a large degree of anti-correlation in the potential arising from a strong spatial correlation among the charged impurities in the substrate.

In this letter we evaluate the ACF of the electrostatic potential in graphene
by using Green's function (GF) for the Poisson equation for a layered structure of dielectrics surrounding graphene,
\cite{Ong_2012,Miskovic_2012,Anicic_2013}
which is combined in a self-consistent manner with the polarization function of graphene within the random phase approximation, where graphene is treated as a zero-thickness material.\cite{Sarma_2011}
Specifically, we explore the effects of finite correlation length $r_c$ between point-charge impurities distributed in a two-dimensional (2D) layer parallel to graphene,\cite{Yan_2011,Li_2011} as well as the effects of finite thickness of a high-$\kappa$ dielectric layer used in the configuration of a top-gated graphene.\cite{Ong_2012,Fallahazad_2010,Hollander_2011}
We show that the experimentally observed negative values in the ACF of the potential may be modeled by invoking sufficiently large correlation lengths, on the order of $r_c\sim$ 10 nm.
\cite{Zhang_2009,Deshpande_2011,Gomez_2012}
However, given the typically large equilibrium charge carrier density of graphene on SiO$_2$, on the order of $\barn\sim 10^{12}$ cm$^{-2}$, one expects that equally large areal density of charged impurities, $\nimp$, would yield a typically quite large packing fraction, $p=\pi \nimp r_c^2/4$, in a 2D distribution of charged impurities.
Therefore, statistical description of the charged impurities must go beyond the use of a simple step-like pair correlation function that is valid for a 2D gas-like structure with $p\ll 1$.\cite{Yan_2011,Li_2011} For that purpose we use here an analytically parameterized model of hard disks (HD) due to Rosenfeld,\cite{Rosenfeld_1990,Anicic_2013} which gives reliable results for packing fractions up to the freezing point of a 2D fluid, $p\approx$ 0.69.

Using a three-dimensional (3D) Cartesian coordinate system with coordinates $\{x,y,z\}$,
we assume that a single-layer graphene sheet of large area $A$ is placed in the plane $z=z_g$ and is embedded into a stratified structure of dielectric layers with the interfaces parallel to graphene.
Assuming that the entire structure is translationally invariant (and isotropic) in the directions of a 2D position vector $\br=\{x,y\}$, we perform a 2D Fourier transform (FT) ($\br\rightarrow\bq$) of the GF for the entire structure \emph{without} graphene, $G(\br-\br';z,z')$, the fluctuation of the electrostatic potential in graphene, $\Dphi(\br)$, and the fluctuation in the density of external charged impurities, $\Drho(\br,z)$. Denoting all the FTs with a tilde, one may write\cite{Miskovic_2012,Anicic_2013}
\begin{eqnarray}
\DFphi(\bq)&=&\int\limits dz\, \frac{\FG(q;z_g,z)\,\DFrho(\bq,z)}{1+e^2\chi(q)\FG(q;z_g,z_g)},
\end{eqnarray}
where $e$ is the elementary charge and $\chi(q)$ is the static polarization function of graphene,\cite{Wunsch_2006,Hwang_2007} while the expressions for the FT of the GF (FTGF) for a three-layer structure of dielectrics are given elsewhere.\cite{Miskovic_2012,Anicic_2013} We note that the screened Coulomb interactions due to nearby dielectrics is governed by the FTGF $\FG(q;z,z')$, while the screening by graphene is assumed to be unaffected by the fluctuation in the equilibrium charge carrier density across graphene and is adequately described by the function $\chi(q)$ that depends on $\barn$.

Assuming that $N$ impurities are randomly distributed in the dielectric under the area $A$ covered by graphene, we seek an ensemble average, denoted by $\langle\cdots\rangle$, that defines the ACF of the potential in the plane of graphene as $\langle\Dphi(\br)\,\Dphi(\br')\rangle\equiv C_V(\br-\br')$. We obtain from the inverse FT of Eq.~(1)
\begin{eqnarray}
C_V(\br)=e^2\nimp\!\int\!\frac{d^2\bq}{(2\pi)^2}\,\frac{\mbox{e}^{i\bq\cdot\br}}{\ve^2(q)}\,S(q),
\end{eqnarray}
where $\nimp=N/A$ is the average areal density of impurities, $\ve(q)\equiv 1+e^2\chi(q)\FG(q;z_g,z_g)$ is the effective dielectric function of graphene, whereas the Coulomb structure factor of the impurities is given by
\begin{eqnarray}
S(\bq)&=&\!\int\! dz\,f(z)\FG^2(q;z_g,z)+\nimp\!\int\! dz\,f(z)\FG(q;z_g,z)
\nonumber\\
&\times&\!\!
\!\int\! dz'\,f(z')\FG(q;z_g,z')
\!\int\! d^2\br\,\mathrm{e}^{i\bq\cdot\br}\left[g(\br;z,z')\!-\!1\right]
\end{eqnarray}
with $f(z)$ being the distribution of the impurity positions along the $z$ axis (assumed to be normalized to one),
and $g(\br;z,z')$ the usual pair correlation function. We consider a 2D distribution of impurities placed in the plane $z=z_i$ with $f(z)=\delta(z-z_i)$, having the radial distribution function $g(\br;z,z')=g(r)$ described by two models that contain $r_c$ as single parameter: a step-correlation (SC) model with $g(r)=1$ for $r>r_c$ and $g(r)=0$ otherwise, which was often used in the previous studies of charged impurities in graphene,\cite{Yan_2011,Li_2011} and the HD model, in which particles interact with each other as hard disks of the diameter $r_c$. \cite{Rosenfeld_1990} In the case of a top-gated graphene we also consider a 3D distribution of uncorrelated impurities with $g(\br;z,z')=1$, which are homogeneously distributed throughout a dielectric slab of finite thickness $L$, so that $f(z)=1/(L-d)$, where $d$ is the minimum distance of impurities from graphene.

\begin{figure}
\centering
\includegraphics[width=0.4\textwidth]{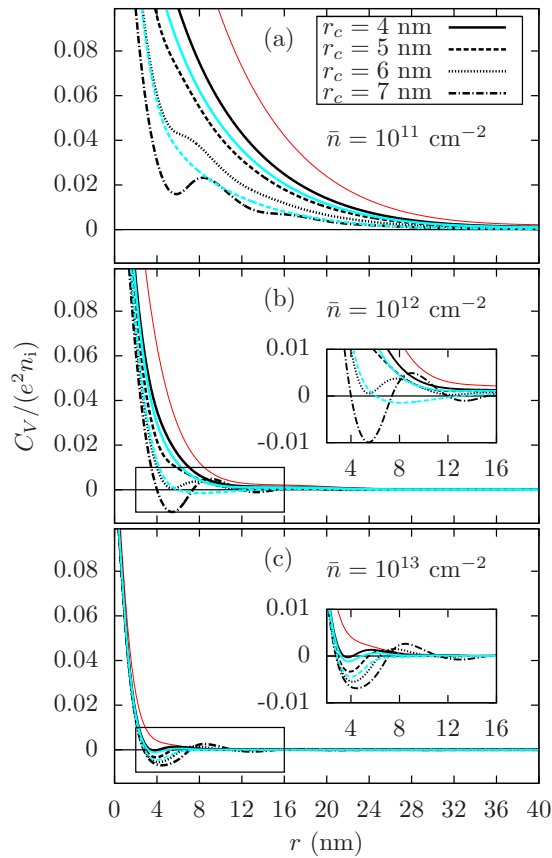}
\caption{(Color online)
The dependence of autocorrelation function of the potential (normalized by $e^2\nimp$) on distance $r$ (in nm) for graphene laying at the boundary between a semi-infinite SiO$_2$ substrate and air, with three charge carrier densities in graphene: (a) $\barn=10^{11}$,  (b)  $\barn=10^{12}$, and (c) $\barn=10^{13}$ cm$^{-2}$. A planar distribution of charged impurities with the number density $\nimp=10^{12}$ cm$^{-2}$ is placed in SiO$_2$ at a depth $d=$ 0.3 nm, having the correlation length $r_c$.
Results are shown for uncorrelated impurities [thin gray (red) solid lines], for the SC model with $r_c=$ 4 and 5 nm [thick solid and dashed gray (cyan) lines, respectively], and for the HD model with $r_c=$ 4, 5, 6, and 7 nm (thick black solid, dashed, dotted, and dash-dotted lines, respectively). Insets show enlarged regions with 2 nm $<r<$ 16 nm.}
\end{figure}
In Fig.~1 we consider a two-layer structure that consists of semi-infinite regions of SiO$_2$ and air with graphene placed on the surface of SiO$_2$, and a planar layer of point-charge impurities embedded in SiO$_2$ a distance $d=0.3$ nm away from graphene. We show the radial dependence of the ACF for several values of $\barn$ and for
different correlation lengths $r_c$ that are treated by both the SC and the HD models. One notices in Fig.~1 that the overall range of the ACF decreases with increasing $\barn$ owing to the screening by graphene, which is characterized by the screening length $\propto k_F^{-1}$ where $k_F=\sqrt{\pi\barn}$ is the Fermi wavenumber of graphene.
More importantly, one notices in Fig.~1 that oscillations develop in the ACF for increasing $r_c$ and $\barn$ values, with no oscillations ever observed for the uncorrelated impurities with $r_c=0$. For sufficiently large $r_c$ and $\barn$ values both the SC and the HD models give rise to negative values in the ACF over certain intervals of the radial distance $r$. While the SC model gives more articulate oscillations than the HD model for the same value of $r_c$, its domain of applicability is limited to correlation lengths $r_c<5.6$ nm for the given value of $\nimp=10^{12}$ cm$^{-2}$ because the SC model breaks down for $p>0.25$. On the other hand, the HD model allows the use of much larger $r_c$ values than the SC model, giving rise to stronger oscillations in the ACF than those that could be achieved with the SC model.
Remarkably, for the highest value of $\barn$ shown in Fig.~1(c), one notices that the second zero in the ACF in the HD model occurs at a distance that is approximately equal to the corresponding correlation length $r_c$.
This may be rationalized by noticing that, due to heavy screening by graphene at large $\barn$, the
fluctuations in both the electrostatic potential and the charge carrier density in graphene resemble the geometric structure of the underlying charged impurities represented by hard discs with the diameter $r_c$. In view of this finding it is interesting to note that the values of the second zero in the ACF found in various experiments are quite large: 30 nm \cite{Zhang_2009}, 65-90 nm \cite{Deshpande_2011}, and 18 nm. \cite{Gomez_2012}

\begin{figure}
\centering
\includegraphics[width=0.45\textwidth]{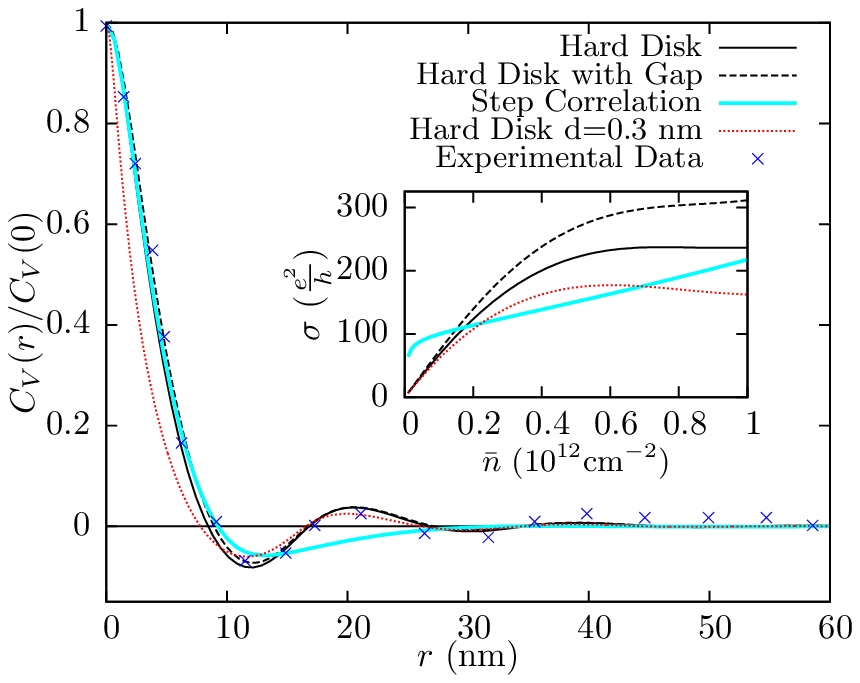}
\caption{(Color online)
The dependence of the normalized autocorrelation function of the potential $C_V(r)/C_V(0)$ on distance $r$ (in nm) for graphene laying at the boundary between a semi-infinite SiO$_2$ substrate and air, with the charge carrier density in graphene $\barn=2.5\times 10^{11}$ cm$^{-2}$ and a planar distribution of charged impurities with the number density $\nimp=2.1\times 10^{11}$ cm$^{-2}$ placed in SiO$_2$ a distance $d$ from graphene. Parameters are chosen to fit the experimental data (symbols) of Ref.\cite{Gomez_2012} by using $d=$ 1 nm for both the HD model with the correlation length $r_c=16$ nm (solid black lines) and the SC model with $r_c=$ 12 nm [thick solid gray (cyan) lines].
Also shown are the results for the HD model with the same parameters, but with the air gap of 0.3 nm between graphene and SiO$_2$ (dashed black lines), and for a reduced distance of $d=$ 0.3 nm with zero gap [thin dotted gray (red) lines].
The inset shows the conductivity $\sigma$ of graphene (in units $e^2/h$) as a function of $\barn$ (in units $10^{12}$ cm$^{-2}$).
 }
\end{figure}

\begin{figure}
\centering
\includegraphics[width=0.4\textwidth]{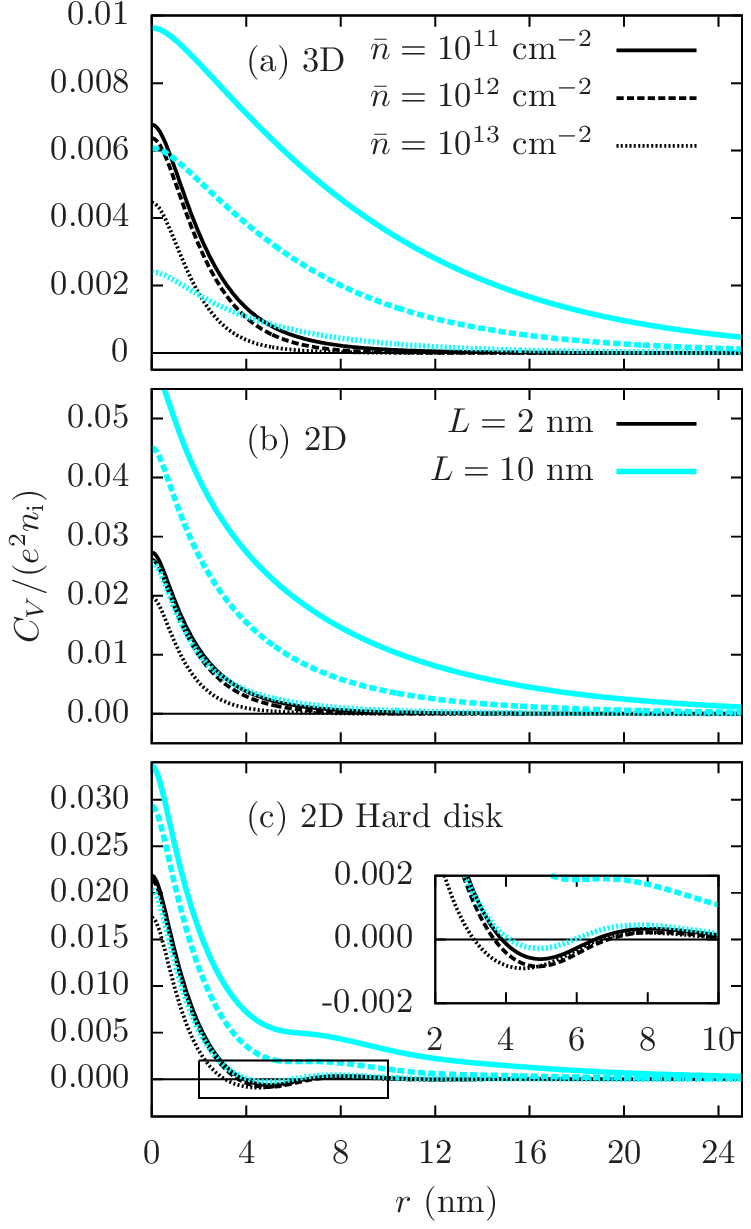}
\caption{(Color online)
The dependence of autocorrelation function of the potential (normalized by $e^2\nimp$) on distance $r$ (in nm) for graphene laying at the boundary between a semi-infinite SiO$_2$ substrate and a HfO$_2$ layer of thickness $L=$ 2 nm (thin black lines) and 10 nm [thick gray (cyan) lines], with the charge carrier density in graphene $\barn=10^{11}$ (solid lines), $\barn=10^{12}$ (dashed lines), and $\barn=10^{13}$ cm$^{-2}$ (dotted lines). Results are shown for (a) a 3D homogeneous distribution of impurities throughout the HfO$_2$ layer extending up to a distance $d=$ 0.3 nm from graphene, as well as for a planar 2D distribution of impurities placed in the HfO$_2$ layer a distance $d=$ 0.3 nm from graphene for both (b) uncorrelated impurities and (c) the correlation length $r_c=$ 6 nm treated by the HD model.
 }
\end{figure}
In Fig.~2 we attempt to model the normalized ACF of the potential, $C_V(r)/C_V(0)$, deduced from the experimental data for graphene on SiO$_2$, \cite{Gomez_2012}
by using $r_c$ and the distance $d$ of impurities from graphene as fitting parameters, while assigning to $\nimp$ and $\barn$ the values commensurate with those found in the experiment. We find the best fit to the experiment by
using the HD model with $r_c=16$ nm for $\nimp=2.1\times 10^{11}$ cm$^{-2}$ (giving $p=0.42$) and $\barn=2.5\times 10^{11}$ cm$^{-2}$,
while setting $d=1$ nm, which is commensurate with distances used in modeling of the conductivity of graphene on SiO$_2$. \cite{PNAS_2007,Sarma_2011}
We also show in Fig,~2 two other results for the HD model with the same $r_c$, $\nimp$ and $\barn$ values, where we reduce the distance to $d=$ 0.3 nm,
or allow for a gap of air of thickness 0.3 nm between graphene and the SiO$_2$ surface while keeping $d=$ 1 nm. \cite{Ishigami_2007,Romero_2008}
Also shown is the best fit with the SC model that is achieved for $r_c=12$ nm (giving $p=0.24$) with all other parameters having the same values as those used for the best fit with the HD model.

In the inset to Fig.~2 we show the conductivity $\sigma$ as a function of $\barn$ for graphene on SiO$_2$ evaluated from the semiclassical Boltzmann transport theory for the four cases discussed in the main panel.\cite{PNAS_2007,Sarma_2011,Anicic_2013}
One notices a linear increase of $\sigma$ at small $\barn$ (with a much larger slope for the SC model than for the HD model), followed by a sublinear behavior of $\sigma $ at large $\barn$ values, which is represented by a reduction in the slope of $\sigma$ for the SC model and a saturation in $\sigma$ for the HD model.\cite{Anicic_2013}

One notices in Fig.~2 that the existence of a finite gap of air has negligible effect on the ACF in the HD model, but noticeably reduces the saturation rate in the conductivity at high $\barn$ values. On the other hand, a reduction of the distance $d$ of impurities with zero air gap reduces the width of the main peak in the normalized ACF, which extends for distances from $r=0$ to the first zero in ACF, and accentuates the saturation in the conductivity in the HD model at high $\barn$ values.
Most importantly, besides reproducing the main peak and the first zero of the normalized ACF,
the SC model fails to reproduce the shape of the experimental ACF including its second zero and the subsequent peaks and valleys, which are well reproduced by the HD model. We note that our using the packing fraction of $p=0.24$ stretches the SC model to its breakdown point, whereas reduction in the packing fraction would further worsen the agreement of the SC model with the experiment.
Therefore, one may conclude that the structure of the normalized ACF in this example is a result of a rather strong correlation among the charged impurities with the correlation length of $r_c=16$ nm that gives rise to a large packing fraction $p=0.42$, which may be reliably described by the HD model.

It should come as no surprise that the correlation lengths among charged impurities may reach such large values in the presence of graphene. It was recently shown that the interaction potential between two point charges near doped graphene is heavily screened and, moreover, exhibits Friedel oscillations with inter-particle distance, giving rise to a repulsive core region of distances on the order of $k_F^{-1}$ that resembles the interaction among hard disks with diameter $r_c\sim k_F^{-1}$.\cite{Radovic_2012}

In Fig.~3 we consider single layer graphene sandwiched between a semi-infinite layer of SiO$_2$ and a layer of HfO$_2$ of finite thickness $L$, which is typical for top-gating through a high-$\kappa$ dielectric.\cite{Ong_2012,Fallahazad_2010,Hollander_2011} We show the radial dependence of the ACF for several combinations of the $\barn$ and $L$ values, and for several model distributions of point-charge impurities in the HfO$_2$ layer having the areal number density $\nimp=10^{12}$ cm$^{-2}$. We consider a homogeneous 3D distribution of uncorrelated charges throughout the HfO$_2$, which extends up to a distance $d=$ 0.3 nm from graphene, as well as a 2D planar distribution placed in HfO$_2$ a distance $d=$ 0.3 nm away from graphene, with both uncorrelated ($r_c=0$) and correlated ($r_c=6$ nm, $p\approx 0.28$) charges that are described with the HD model.
In comparison to Fig.~1, one notices that ACF has generally smaller magnitude in Fig.~3 because of stronger screening due to much larger dielectric constants involved.
Moreover, comparing various cases of the distribution of impurities in Fig.~3, one notices that the ACF has a much lower magnitude in the case of a 3D distribution than in the corresponding 2D cases because the impurities are spread over larger distances from graphene in the 3D case and hence the resulting potential and its fluctuations are weaker.
One further notices in Fig.~3 that a reduction in thickness $L$ suppresses the overall range of the ACF in a similar manner as does the increase in $\barn$, which is caused by the increased screening due to the proximity of a perfectly conducting gate on the opposite boundary of the HfO$_2$ layer from graphene. This increased screening by the gate is also responsible for the more articulated oscillations in the ACF for smaller thicknesses $L$.

In conclusion, we have shown that both an increase of charge carrier density in graphene and a reduction of the distance of a nearby gate provide strong screening effects in the autocorrelation function (ACF) of the electrostatic potential in graphene. Those effects help reveal spatial correlation between charged impurities in the dielectric through appearance of oscillations of the ACF as a function of distance that exhibit well-defined intervals of anti-correlation in the potential. We have found that the second zero in the ACF is related to the correlation length for a 2D distribution of impurities, which may take quite large values, according to several experiments. Consequently, statistical models for the structure of charged impurities near graphene must be able to tackle systems with large packing fractions, resembling charged 2D fluids.

\begin{acknowledgments}
This work was supported by the Natural Sciences and Engineering Research Council of Canada.
\end{acknowledgments}

\newpage

\newpage

\newpage


\begin{thebibliography}{10}

\bibitem{Avouris_2012}
Ph. Avouris and F. Xia, MRS Bulletin \textbf{37}, 1225 (2012).

\bibitem{Allen_2010} M. J. Allen, V. C. Tung, and R. B. Kaner, Chem. Rev. \textbf{110},132 (2010).

\bibitem{Ishigami_2007}
M. Ishigami, J. H. Chen, W. G. Cullen, M. S. Fuhrer, and E. D. Williams, Nano Lett. \textbf{7}, 1643 (2007).

\bibitem{Romero_2008}
H. Romero, N. Shen, P. Joshi, H. R. Gutierrez, S. A. Tadigadapa, J. O. Sofo, and P. C. Eklund, ACS Nano 2, 2037 (2008).

\bibitem{Martin_2008}
J. Martin, N. Akerman, G. Ulbricht, T. Lohmann, J.H. Smet, K. Von Klitzing, and A. Jacoby, Nat. Phys. \textbf{4}, 144 (2008).

\bibitem{Zhang_2009}
Y. Zhang, V.W. Brar, C. Girit, A. Zettl, M.F. Crommie, Nat. Phys. \textbf{5},  722 (2009).

\bibitem{Deshpande_2009}
A. Deshpande, W. Bao, Z. Zhao, C. N. Lau, and B. J. LeRoy, Appl. Phys. Lett. 95, 243502 (2009).

\bibitem{Deshpande_2011}
A. Deshpande, W. Bao, Z. Zhao, C. N. Lau, and B. J. LeRoy, Phys. Rev. B 83, 155409 (2011).

\bibitem{Gomez_2012}
A. Castellanos-Gomez, R. H. M. Smit, N. Agrait, and G. Rubio-Bollinger, Carbon 50, 932 (2012).

\bibitem{Tan_2007}
Y.-W. Tan, Y. Zhang, K. Bolotin, Y. Zhao, S. Adam, E. H. Hwang, S. Das Sarma, H. L. Stormer, and P. Kim, Phys. Rev. Lett. 99, 246803 (2007).

\bibitem{PNAS_2007}
S. Adam, E. H. Hwang, V. M. Galitskii, and S. Das Sarma, Proc. Natl. Acad. USA \textbf{104}, 18392 (2007).

\bibitem{Sarma_2011}
S. Das Sarma, S. Adam, E. H. Hwang, and E. Rossi, Rev. Mod. Phys. \textbf{83}, 407 (2011).

\bibitem{Yan_2011}
J. Yan and M. S. Fuhrer, Phys. Rev. Lett. \textbf{107}, 206601 (2011).

\bibitem{Li_2011}
Q. Li, E. H. Hwang, E. Rossi, and S. Das Sarma, Phys. Rev. Lett. \textbf{107}, 156601 (2011).

\bibitem{Yan_2013}
H. Yan, T. Low, W. Zhu, Y. Wu, M. Freitag, X. Li, F. Guinea, P. Avouris, and F. Xia, Nat. Photonics \textbf{7}, 394 (2013).

\bibitem{Ong_2012}
Z.-Y. Ong and M. V. Fischetti,  Phys. Rev. B \textbf{86}, 121409(R) (2012). 

\bibitem{Miskovic_2012}
Z.L. Miskovic, P. Sharma and F. O. Goodman, Phys. Rev. B \textbf{86}, 115437 (2012).

\bibitem{Anicic_2013}
R. Anicic and Z. L. Miskovic,
arXiv:1307.8169 (2013).

\bibitem{Fallahazad_2010}
B. Fallahazad, S. Kim, L. Colombo, and E. Tutuc, Appl. Phys. Lett. \textbf{97}, 123105 (2010).


\bibitem{Hollander_2011}
M. J. Hollander, M. LaBella, Z. R. Hughes, M. Zhu, K. A. Trumbull, R. Cavalero,  D. W. Snyder, X. Wang, E. Hwang, S. Datta, and J. A. Robinson, Nano Lett. \textbf{11}, 3601 (2011).

\bibitem{Rosenfeld_1990}
Y. Rosenfeld,  Phys. Rev. A \textbf{42}, 5978 (1990).

\bibitem{Wunsch_2006}
B. Wunsch, T. Stauber, F. Sols, and F. Guinea, New J. Phys. \textbf{8}, 318 (2006).

\bibitem{Hwang_2007}
E. H. Hwang and S. Das Sarma, Phys. Rev. B \textbf{75}, 205418 (2007).

\bibitem{Radovic_2012}
I. Radovic, D. Borka, and Z. L. Miskovic, Phys. Rev. B \textbf{86}, 125442 (2012).


\end{thebibliography}
\end{document}